# The sum of it all: revealing collaboration patterns by combining authorship and acknowledgements

Adèle Paul-Hus[1], Philippe Mongeon[1], Maxime Sainte-Marie[1], Vincent Larivière[2]

[1]adele.paul-hus@umontreal.ca, philippe.mongeon@umontreal.ca, maxime.sainte-marie@umontreal.ca
École de bibliothéconomie et des sciences de l'information, Université de Montréal
PO Box 6128, Downtown Station, Montreal, Quebec, H3C 3J7 (Canada)

[2] vincent.lariviere@umontreal.ca
École de bibliothéconomie et des sciences de l'information, Université de Montréal
PO Box 6128, Downtown Station, Montreal, Quebec, H3C 3J7 (Canada) and
Observatoire des Sciences et des Technologies (OST), Centre Interuniversitaire de Recherche sur la Science et la Technologie (CIRST), Université du Québec à Montréal,
PO Box 8888, Downtown Station, H3C 3P8 Montréal, Qc. (Canada)

**Abstract**

Acknowledgments are one of many conventions by which researchers publicly bestow recognition towards individuals, organizations and institutions that contributed in some way to the work that led to publication. Combining data on both co-authors and acknowledged individuals, the present study analyses disciplinary differences in researchers' credit attribution practices in collaborative context. Our results show that the important differences traditionally observed between disciplines in terms of team size are greatly reduced when acknowledgees are taken into account. Broadening the measurement of collaboration beyond co-authorship by including individuals credited in the acknowledgements allows for an assessment of collaboration practices and team work that might be closer to the reality of contemporary research, especially in the social sciences and humanities.

**1. Introduction**

Acknowledgments are one of many conventions by which researchers give credit and publicly share gratitude and recognition towards individuals, organizations and institutions that contributed to the work that led to publication. Although they could be perceived as the "scholar's courtesy" (Cronin, 1995), acknowledgements convey rich information that can shed light on researchers' collaborative activities that cannot be revealed by analysing co-authorship. In that sense, acknowledgements can be conceived as markers of symbolic capital (Bourdieu, 1975) that complements authorship, and have been included as a component of the "reward triangle" alongside authorships and citations (Cronin & Weaver-Wozniak, 1993). In most natural and biomedical sciences disciplines, teamwork constitutes the norm rather than the exception (Cronin, 2004; Wuchty, Jones & Uzzi, 2007). Henriksen (2016) and Larivière, Gingras and Archambault (2006) have further shown that the rise in research collaborations also extends to most social sciences disciplines, in terms of average number of authors, share of co-authored articles, as well as international collaboration. However, these results, as most bibliometric investigations of collaboration, are limited to formal collaborations as measured by co-authorship. Indeed, as highlighted by Katz and Martin (1997), many instances of collaboration do not lead to co-authorship, while indirect interactions between researchers might actually do. This has led them to conclude that co-authorship is a "rather imperfect or partial indicator of research collaboration between individuals." Katz and Martin (1997, p. 11).

Laudel (2002) also challenged that traditional bibliometric practice of using co-authorships as a proxy for research collaboration and identified six types of research collaborations associated to distinct patterns of rewards. Based on interviews with researchers and an analysis of 133 publications, Laudel (2002) showed that, while some contributions were associated with authorship, one third of all contributions analysed were only rewarded by acknowledgements and about half of contributions were not associated to any public recognition and were thus invisible in formal communication channels. More recently, Ponomariov and Boardman (2016) surveyed academic researchers on their relationship with their collaborators and showed that in many instances, collaboration does not entail co-authorship, a finding which leads the authors to suggest using data that go beyond co-authorship when studying collaboration.

Types of contributions that get rewarded by authorship vary in their nature but also by field, discipline and specific teamwork culture (Larivière et al., 2016). High Energy Physics (HEP) represents a telling example of discipline-specific authorship attribution practices, with projects typically involving thousands of individuals and almost as many institutions. In that context, specific guidelines govern authorship. For instance, all members the project are included in a standard author list and each paper emerging from the project will be alphabetically co-authored by all those on the list (Biagioli, 2004, Birnholtz, 2006). In 2015, a new record for the largest number of authors on a single research article has been set by a HEP publication, co-signed by more 5,000 individuals (Castelvecchi, 2015). A contrasting example is found in medical research, where the notion of authorship is closely linked to responsibility and accountability. Given the dangerous consequences associated to fraud in those disciplines and its rising co-authorship rates, the International

Committee of Medical Journal Editors (ICMJE) published, for the first time in 1988, the *Recommendations for the Conduct, Reporting, Editing, and Publication of Scholarly Work in Medical Journals*. Updated in 2015, the ICMJE criteria recommends that authorship be based on:

- substantial contributions to the conception or design of the work; or the acquisition, analysis, or interpretation of data for the work, AND
- drafting the work or revising it critically for important intellectual content, AND
- final approval of the version to be published, AND
- agreement to be accountable for all aspects of the work in ensuring that questions related to the accuracy or integrity of any part of the work are appropriately investigated and resolved. (ICMJE, 2015: 2).

Moreover, "contributors who meet fewer than all 4 of the above criteria for authorship should not be listed as authors, but they should be acknowledged" (ICMJE, 2015:3). This suggests that when the ICMJE guidelines are strictly followed, many contributions may be insufficient to warrant authorship and should rather be rewarded by an acknowledgement only.

Contrasting with the ICMJE authorship guidelines, Rennie, Yank and Emmanuel (1997) proposed that the notion of author is "outmoded", and that it cannot appropriately account for credit and responsibility in multi-authors publications[1]. They proposed a system where the notion of contributorship would replace the notion of authorship. The main objective of their proposition was to ensure more equitable and reliable credit and responsibility attribution practices, where all collaborators would systematically disclose their specific contributions. This radical alternative would eliminate "the artificial distinction, mostly of a social nature, between authors and non-author contributors—that is, between 'authors' and 'acknowledgees'" (Rennie, Yank & Emmanuel, 1997, p. 584). Almost two decades later, the contributorship model, as envisionned originally, has not been implemented anywhere. However, many journals, mostly in the medical field, now include contribution

---

[1] It should be noted that the ICMJE authorship guidelines were slightly different at the time of Rennie, Yank and Emmanuel proposal and consisted of the following: "Authorship credit should be based only on substantial contributions to (a) conception and design, or analysis and interpretation of data; and to (b) drafting the article or revising it critically for important intellectual content; and on (c) final approval of the version to be published. Conditions (a), (b), and (c) must all be met." (ICMJE, 1997:311).

statements (e.g.Nature, PNAS, the British Medical Journal and the PLOS series of journals).

Notwithstanding their potential to reveal often invisible contributions to research, the current format of acknowledgements limits their use. As highlighted by McCain (1991), "[t]he format of acknowledgment varies from field to field and from journal to journal. As noted, persons and institutional sources may be listed in the methods and materials section of an article or explicitly thanked in an acknowledgements section" (p.506). This lack of standardization—highlighted by many researchers (e.g. Cronin, 1995; Paisley & Parker, 1967; Mackintosh, 1972; Giles & Council, 2004)—has contributed to the ambiguous reputation of acknowledgements in the scientific community. This unstandardized space of thanking leads to very heterogeneous testimonies of gratitude, and contributions getting rewarded by an acknowledgement can be even more heterogeneous than those leading to authorship. On the one hand, Cronin, McKenzie, Rubio and Weaver-Wozniack's (1993) classification of acknowledgements ranges from conceptual and intellectual contributions to provision of financial support, access to data and materials, technical assistance and manuscript preparation; these same types of contributions can be sufficient to warrant authorship in certain contexts. On the other hand, contributions that could be perceived as trivial or hardly relevant in light of most authorship criteria can lead to authorship in some instances. For example, in a recent article, one of the authors' contribution consisted in driving the car during the data collection process[2]. Similarly, several studies have reported a high prevalence of honorific (or gift) authorship, where researchers who did not make a substantial contribution to a work (or did not contribute at all in some cases), but are included in the author list (e.g. Flanagin et al., 1998; Marušić, Bošnjak & Jerončić, 2011 Wislar, Flanagin, Fontanarosa & DeAngelis, 2011). This diversity of disciplinary, but also individual, authorship attribution practices—some of which being more inclusive than others—can induce artificial distinctions in team size and collaboration, as measured by co-authorship. This highlights the need for new methods that transcend such limitations and provide a more accurate assessment of collaboration in research. This paper attempts to do so by combining acknowledgements and authorship data in order to explore the potential of acknowledgements to reveal collaboration practices going beyond authorship analysis

---

[2] https://twitter.com/igoodfel/status/732927411650744320

Since Giles and Council (2004) pioneering analysis of more than 180,000 acknowledgements found in computer science papers, no large-scale investigation of individuals acknowledged were performed. The present study aims at filling this gap, by analysing more than 1,000,000 scholarly documents containing acknowledgements. In order to extend the notion of collaboration beyond authorship, we analyse formal and informal collaborations. Our unit of analysis thus includes broader types of contributions that are credited by authorship in certain contexts and not in others, when they are made visible in acknowledgements. The individuals involved in a research paper, and credited for it (either formally by authorship or informally by a mention in the acknowledgements) will thus for the purpose of this study be designated as contributors. Collaboration is hence defined inclusively and operationalized as papers having at least two contributors credited on a paper, mentioned either in the byline or the acknowledgements text. The objective of this study is to compare the credit attribution practices of researchers in natural, medical and social sciences. More specifically, we aim at answering the following research questions:

- How many contributors are credited on scholarly publications and how does this vary by discipline?
- What share of contributors is credited as authors and what share is credited as acknowledgees and how does it vary by discipline?
- How does the number of acknowledgees vary as a function of the number of authors signing a scholarly publication and how does this relationship vary by discipline?

**2. Material and methods**

2.1 Data

Data for this study were drawn from Web of Science (WoS) Science Citation Index Expanded (SCI-E) and Social Sciences Citation Index (SSCI), which include acknowledgement data. These acknowledgement data are structured in three fields: the 'Funding Text' (FT), 'Funding Agency' (FO) and 'Grant Number' (FG). FT is the full text of acknowledgements, as it appears in the paper from which it is retrieved. However, as shown by Paul-Hus, Desrochers and Costas (2016), acknowledgements texts are collected and indexed by WoS only if they include funding information. The sum of contributors (authors and acknowledgees) here analysed is consequently limited to publications where a source of funding is acknowledged.

Although WoS started the collection of acknowledgements data in August 2008 for SCI-E articles and reviews, the collection of these data only started in 2015 for SSCI publications (Paul-Hus, Desrochers & Costas, 2016). A dataset of acknowledgement texts was derived from all 2015 articles and reviews from all disciplines covered by SCI-E and SSCI: Biology, Biomedical Research, Chemistry, Clinical Medicine, Earth and Space, Engineering and Technology, Health, Mathematics, Physics, Professional Fields, Psychology and Social Sciences. The dataset includes a total of 1,009,411 papers with acknowledgements texts, which corresponds to 67.1% of all articles and reviews published in 2015 (Table 1). Discipline assignation was done using the NSF field classification of journals (National Science Foundation, 2006); since the NSF classification assigns only one discipline specialty to each journal, this prevents the double counting of papers.

2.2 Analysis

In order to obtain the number of individuals acknowledged per paper, the Stanford Named Entity Recognizer (NER) (Finkel et al., 2005) module of the Natural Language ToolKit (NLTK) (Bird, 2009) was used on each string of acknowledgment text retrieved from the FT field. Application of the Stanford NER algorithm and selection of all named entities tagged as 'person', led to the extraction of 817,125 distinct person names.

The list of named entities extracted from the acknowledgements was then cleaned in order to eliminate non-human entities. This was done in several steps: incomplete names were first removed from the list (entities containing only a first or last name, or only initials), retaining only entities composed of at least one initial and one last name. In order to remove names not designating actual persons, the list was compared to the list of last names of all authors appearing on publications from 1900 to 2016 in WoS indexes, which includes 2,649,212 distinct last names. This WoS authors list was thus used here as a person-name benchmark list. Entities with no match in this list were considered as not referring to actual individuals and were removed. A further manual cleaning step was done to remove all remaining names that did not refer to individual persons such as grant, foundation, organization and institution names. Examples of such names removed by manual cleaning include: Frederick Banting (grant), Marie Curie (grant and foundation), Boehringer Ingelheim (organization) and Instituto de Salud Carlos III (institution). Finally, acknowledgements often contain the name(s) of the author(s) signing the paper from which the acknowledgements were retrieved. When the name(s) extracted from the acknowledgements of a paper X matched the name of one of the author appearing in the byline of that paper X (using the first initial and the last name), this name was removed from the acknowledgees list for that specific paper, such as in the example below:

> Paper X
> Authors: J. Zhang, X. Feng and Y. Xu
> Acknowledgements text: "Jinsong Zhang, Xiao Feng, and Yong Xu contributed equally to this work […]."

The final list of acknowledgments extracted names includes 810,525 distinct names appearing in 362,767 papers.

## 3. Results

Table 1 presents, by discipline, the number of 2015 articles and reviews, the number (and percentage) of those with acknowledgements, and the number (and percentage) of those that contain at least one acknowledged individual. The proportion of papers in which the acknowledgements include the mention of individuals ranges from 12% (Professional Fields) to 45% (Earth and Space), with an average of 24% all disciplines considered. However, the size of disciplines, in terms of absolute number of papers, varies greatly. Indeed, only 17% of Clinical Medicine papers and 31% of Biomedical Research papers include acknowledgements of specific persons, yet they have the highest number of papers, with respectively 67,019 and 59,142 papers that include the mention of specific individuals in their acknowledgements. The following analysis will focus on the subset of papers that includes funding acknowledgements indexed in WoS, for a total of 1,009,411 papers.

Table 1. Number of 2015 papers, number (and percentage) of papers with acknowledgements, and number (and percentage) of papers with acknowledgees

| Discipline | All papers | Papers with acknowledgements | | Papers with acknowledgees | | |
|---|---|---|---|---|---|---|
| | N | N | % | N | %$_{ack.}$ | %$_{total}$ |
| Earth & Space | 92,238 | 72,922 | 79.1 | 41,633 | 57.1 | 45.1 |
| Biology | 105,279 | 76,281 | 72.5 | 43,365 | 56.8 | 41.2 |
| Biomedical Research | 189,066 | 158,067 | 83.6 | 59,142 | 37.4 | 31.3 |
| Physics | 124,556 | 95,676 | 76.8 | 35,063 | 36.6 | 28.2 |
| Psychology | 31,286 | 15,085 | 48.2 | 7,736 | 51.3 | 24.7 |
| Chemistry | 151,947 | 123,806 | 81.5 | 36,583 | 29.5 | 24.1 |
| Social Sciences | 50,420 | 16,972 | 33.7 | 9,291 | 54.7 | 18.4 |
| Engineering & Technology | 241,124 | 165,590 | 68.7 | 43,899 | 26.5 | 18.2 |
| Clinical Medicine | 389,311 | 218,367 | 56.1 | 67,019 | 30.7 | 17.2 |
| Mathematics | 49,997 | 35,390 | 70.8 | 8,314 | 23.5 | 16.6 |
| Health | 37,309 | 18,703 | 50.1 | 5,651 | 30.2 | 15.1 |
| Professional Fields | 41,015 | 12,552 | 30.6 | 5,071 | 40.4 | 12.4 |
| Total | 1,503,548 | 1,009,411 | 67.1 | 362,767 | 35.9 | 24.1 |

Figure 1 presents the cumulative distribution of papers (with acknowledgements) as a function of the number of author(s) they *at least* contain, by discipline. Three groups of disciplines can be distinguished. The first group, which includes Social Sciences, Mathematics and Professional Fields, has the lowest number of authors per paper, with more than 85% of papers having four authors or less, and with proportions of single authored papers ranging from almost 15% (Professional Fields) to more than 25% (Social Sciences). The group composed of Biomedical Research and Clinical Medicine has the highest levels of co-authorship, with 90% of papers having 12 authors or less, and less than 2% having only one author. The remaining disciplines, mostly from the natural sciences, can be found between those two groups.

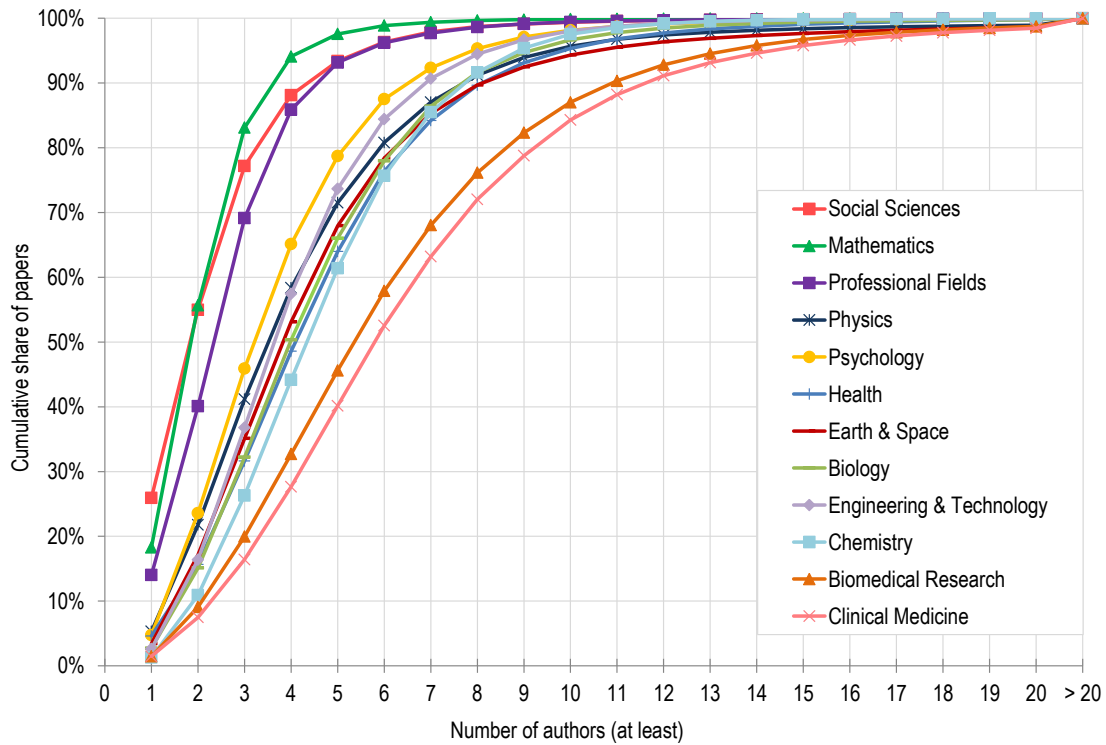

Figure 1. Cumulative distribution of papers (with acknowledgements) (%), as a function of numbers of authors

Figure 2 presents, for all disciplines combined, the distribution of papers (on a log scale) by number of authors (a) and acknowledgees (b), highlighting the skewness of the data. In both cases, the figure clearly shows that the highest proportion of papers is signed by less than 15 authors and acknowledges less than 10 persons. Moreover, both graphs present long-tailed distributions where extreme values of authors and acknowledgees per paper are highly dispersed. Given these data characteristics, the median would generally appear as a more robust measure to describe such distributions. However, because of the high proportion of papers that bears no acknowledgement to specific individuals, the median value of acknowledgees per paper is zero in most disciplines. In this context, the mean value of authors and acknowledgees per paper is deemed the most appropriate measure to describe in a meaningful way the dataset at hand.

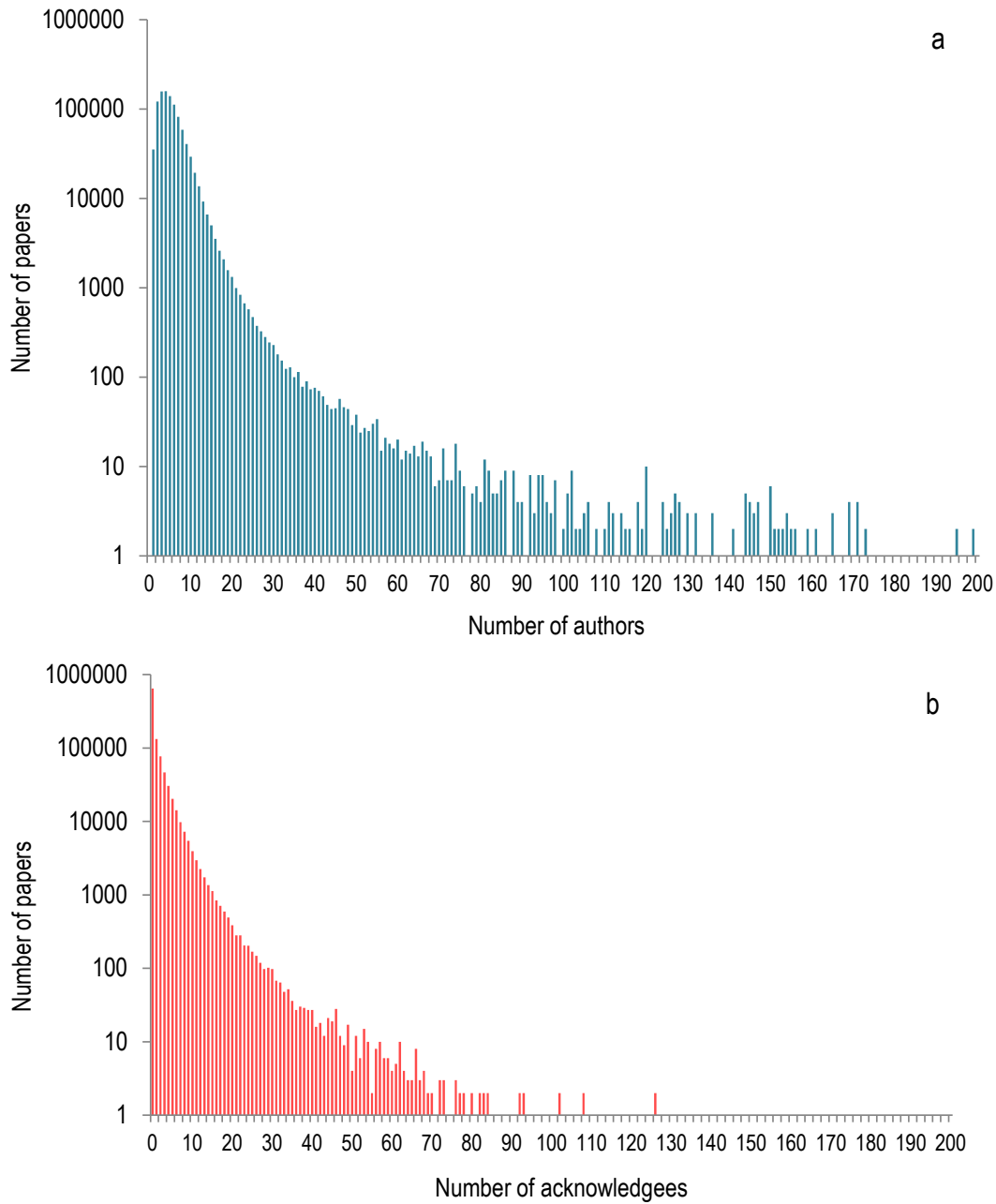

Figure 2. Distribution of papers by number of authors (a) and acknowledgees (b)

Figure 3 presents the mean number of acknowledgees and authors per paper, for the 12 disciplines. The mean number of contributors (the sum of authors and acknowledgees per paper) ranges between 3.1 (Mathematics) and 11.7 (Physics). Figure 3 also displays the variability of the number of contributors per paper with the minimum and maximum number of authors and number of acknowledgees per paper shown in square brackets for each discipline. Physics is by far the discipline where the mean number of contributors is the highest. However, this high number of contributors is mostly attributable to authors, since Physics papers are on average signed by more than 10 authors (10.7) but only acknowledge one person on average. Biomedical Research and Clinical Medicine both have on average more than eight contributors per paper and are similar to Physics in terms of proportion of authors and acknowledgees. In the natural sciences, Earth and Space and Biology differentiate from the other disciplines with about one third of the contributors being acknowledgees. As expected Social Sciences and Professional Fields have low average numbers of authors but, in turn, their average number of acknowledges is similar to that of Earth and Space and Biology. Consequently, these two disciplines stand out with their high acknowledge/author ratio. In Social Sciences, the mean number of acknowledgees per paper (2.8) even exceeds the mean number of authors per paper (2.7).

As a result, disciplines from the social sciences (Health, Psychology, Social Sciences and Professional Fields), which traditionally exhibit much lower level of collaboration when solely considering co-authorship (Larivière, Gingras & Archambault, 2006; Wuchty, Jones & Uzzi, 2007) are displaying mean numbers of contributors that are comparable to what is observed in Chemistry and Engineering and Technology, both natural sciences. Overall, Figure 3 shows that the important differences traditionally observed between all disciplines in terms of team size as measured by co-authorship ($M$ = 4.98, $SD$ = 2.15, $RSD$ = 43%) are greatly reduced when acknowledgees are taken into account ($M$ = 6.68, $SD$ = 2.08, $RSD$ = 31%). This result might indicate that, when considering team size, disciplinary patterns might reflect differences in authorship attribution practices more than actual collaboration practices. In that sense, it suggests that disciplinary differences usually observed in the collaboration level might be amplified by the way we measure collaboration.

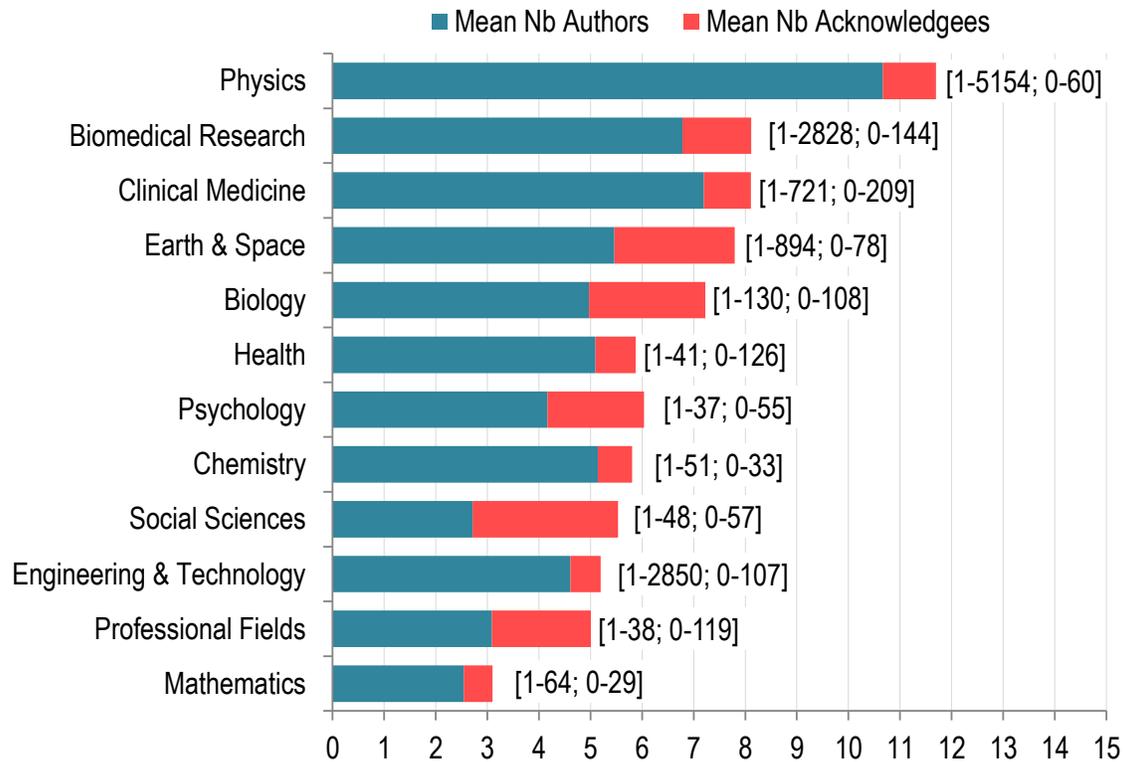

Note: the numbers in the brackets represent the range of the number of author (left) and of the number of acknowledgees (right)

Figure 3. Mean number of authors and acknowledgees, by discipline

For all disciplines, the mean number of acknowledgees decreases or remains stable as the number of authors per paper increases (Figure 4). Moreover, in all disciplines, with the exception of Clinical Medicine and Mathematics, the mean number of acknowledgees is the highest for papers signed by a single author. In Biology, Social Sciences and Professional Fields, a clear decreasing trend is observed in terms of average number of acknowledgees, which implies that as more contributors get credited as authors on collaborative papers, less get acknowledged. For all other disciplines, the mean number of acknowledgees remains stable as the number of authors increases. These trends further support the idea that the lower mean number of authors per paper observed in some disciplines is partly due to less inclusive authorship attribution practices. Mathematics is the exception, standing out as having both the lowest mean number of authors and the lowest mean number of acknowledgees, even for single authored papers.

It should be noted that the relation between the mean number of acknowledgees and the number of authors is presented for values between one and nine on the authors axis since most disciplines have their biggest share papers in that subset, ranging between 100% of papers in Mathematics that have nine authors or less, and 80% of papers in Clinical Medicine that have nine authors or less (see Figure 1 for the complete cumulative distribution). Beyond nine authors, in many disciplines, the number of papers is too small to allow for robust measures.

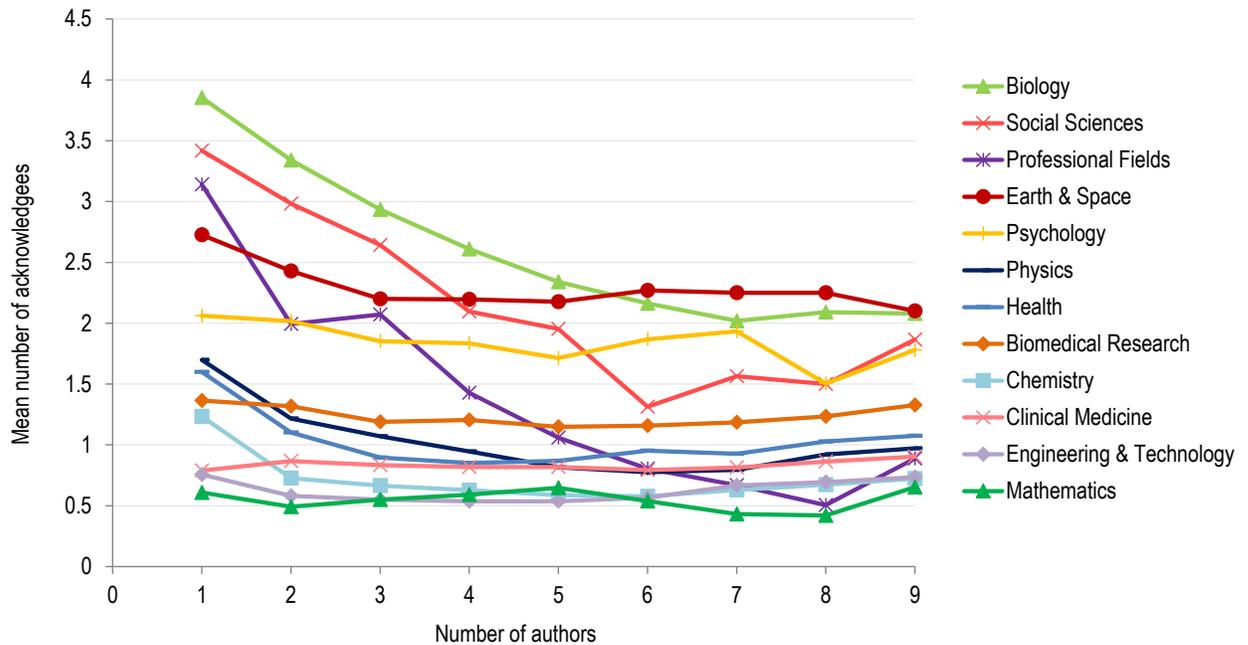

Figure 4. Mean number of acknowledgees by number of authors

## 4. Discussion and conclusion

Over the last decades, scientific collaboration has been the focus of hundreds of bibliometric analyses. However, these analyses have almost all relied on co-authorship as an indicator of collaboration, an operationalization which has been shown to have limitations (Katz & Martin, 1997; Laudel, 2002; Lundberg, Tomson, Lundkvist, Skar & Brommels, 2006). Broadening the measurement of collaboration to include individuals mentioned in the acknowledgements of scholarly publications allows for an assessment of collaboration practices that might be closer to the reality of contemporary research. Our results show that disciplinary differences in collaborative activities are actually much less important, as scholars in the social sciences are collaborating much more than what co-authorship alone suggests. This also confirms that the lone scholar has become an endangered species in most disciplines, including the social sciences. In fact, our data shows that in 40% of publications signed by only one author (restricted to the subset of papers where acknowledgements have been indexed), single authors are not alone since they acknowledge specific individuals who contributed to their research.

One limitation of this study is related to the data source. As mentioned in the methods section, acknowledgements are collected and indexed in WoS only when they contain funding information, thus creating a bias toward funded research projects. Moreover, our analyses are restricted to the Science Citations Index Expanded (SCI-E) and the Social Sciences Citations Index (SSCI) and do not cover publications from Arts and Humanities, which are considered as disciplines with lower collaborative practices (Larivière, Gingras & Archambault, 2006). Yet, WoS still constitutes the most comprehensive source for acknowledgements data. As shown in Table 1, acknowledgements are not evenly distributed among disciplines. Since acknowledgements are not collected systematically, we cannot conclude that acknowledgements are less frequent in those disciplines that exhibit lower shares of funding acknowledgements. We cannot exclude the possibility that a certain number of papers, not analysed in the present study, includes acknowledgements without specifically mentioning funding source. However, our dataset still represents more than two thirds of all articles and reviews published in 2015, a sample size large enough to ensure the robustness of our findings.

Another potential limitation is related to the nature of contributions acknowledged. While most acknowledgements are made to individuals who have actively contributed to the work that led to a publication, there may also be cases where authors acknowledge their partners and family for their support, or other types of "non-scientific" contributions. Nevertheless, our results provide strong evidence of the existence of disciplinary differences not only in terms of authorship practices, but also in terms of acknowledgement practices. Furthermore, these practices appear to influence each other, highlighting the necessity of taking both into account when measuring collaboration in research. However, by assessing the number of individuals involved in the production of scientific publications, we do not weight the value of contributions leading to acknowledgement as equivalent to the ones leading to authorship but rather aim at taking into account the high variability of authorship and acknowledgement practices. In the end, our results suggest that disciplinary differences traditionally observed in terms of team sizes and collaborative activity might be, at least in part, an artifact of the indicator we use to measure collaboration and not a truthful reflection of team size variation between disciplines.

PLOS journals recently introduced a new taxonomy of contributions providing standardized and fine-grained information that "makes transparent who participated and the roles they played" (Atkins, 2016). This new taxonomy has the objective "to know and unambiguously credit [those] who participated in the work being published and forms the base for plans to eventually provide credit to all participants in the research outputs ecosystem" (Atkins, 2016). In order to achieve such unambiguous credit attribution, standardized authorship and acknowledgements criteria would need to be applied and followed rigorously and uniformly across disciplines which, as we have demonstrated in this study, is far from being the case. Moreover, many studies have shown the risks of introducing bias in reporting contributions when authorship and contributorship statements are restricted to a pre-determined taxonomy (Bates, Anić, Marušić & Marušić, 2004; Marušić, Bates, Anić & Marušić, 2006; Ivaniš, Hren, Sambunjak, Marušić & Marušić, 2008). The reliability of such disclosures of contributions tends to be affected by one's own autobiographical memory and perceived value of contribution (Ilakovac, Fister, Marušić & Marušić, 2006; Ivaniš, Hren, Marušić & Marušić, 2011). Nevertheless, the new credit taxonomy from PLOS constitutes a further step towards transparency and accountability for all team members involved in research projects. In turn, a standardized system providing the description of all participants' contributions could lead to a more equitable distribution of credit and reward—which could also contribute to provide a more accurate portrait of what contemporary research involves in terms of humans and materials resources, especially in the social sciences and humanities.

**Acknowledgement**

This research was supported by the Social Sciences and Humanities Research Council of Canada: Joseph-Armand Bombardier CGS Doctoral Scholarships (Paul-Hus, Mongeon); Insight Development Grant (Larivière).

**References**

Atkins, H. (2016). *Author Credit: PLOS and CRediT Update*. The Official PLOS Blog. Retrieved from http://blogs.plos.org/plos/2016/07/author-credit-plos-and-credit-update/

Bates, T., Anić, A., Marusić, M., & Marusić, A. (2004). Authorship criteria and disclosure of contributions: comparison of 3 general medical journals with different author contribution forms. *JAMA*, *292*(1), 86–88. https://doi.org/10.1001/jama.292.1.86

Biagioli, M. (2003). Rights or rewards? Changing frameworks of scientific authorship. In P. Galison & M. Biagioli (Eds.), *Scientific authorship : credit and intellectual property in science* (pp. 255–279). New York, NY: Routledge.

Bird, S., Klein, E., & Loper, E. (2009). Natural Language Processing with Python: Analyzing Text with the Natural Language Toolkit. Sebastopol, CA: O'Reilly Media.

Birnholtz, J. P. (2006). What does it mean to be an author? The intersection of credit, contribution, and collaboration in science. *Journal of the American Society for Information Science and Technology*, *57*(13), 1758–1770. http://doi.org/10.1002/asi.20380

Bourdieu, P. (1975). The specificity of the scientific field and the social conditions of the progress of reason. *Social science information*, *14*(6), 19-47.

Castelvecchi, D. (2015). Physics paper sets record with more than 5,000 authors. *Nature News.* http://doi.org/10.1038/nature.2015.17567

Cronin, B. (2004). Bowling alone together: Academic writing as distributed cognition. *Journal of the American Society for Information Science and Technology*, *55*(6), 557-560. http://doi.org/10.1002/asi.10406

Cronin, B. (1995). *The scholar's courtesy : the role of acknowledgement in the primary communication process.* London: Taylor Graham.

Cronin, B., & Weaver-Wozniak, S. (1993). Online access to acknowledgements. In Martha E. Williams (Ed).*Proceedings of the Fourteenth National Online Meeting 1993.* (p.93-98). Medford: NY, Learned Information, Inc.

Cronin, B., McKenzie, G., Rubio, L., & Weaver-Wozniak, S. (1993). Accounting for influence: Acknowledgments in contemporary sociology. *Journal of the American Society for Information Science*, *44*(7), 406–412.

Finkel, J.R., Grenager T., and Manning, C. (2005*). Incorporating Non-local Information into Information Extraction Systems by Gibbs Sampling.* Proceedings of the 43nd Annual Meeting of the Association for Computational Linguistics (ACL 2005), pp. 363-370.

Flanagin, A., Carey, L. A., Fontanarosa, P. B., Phillips, S. G., Pace, B. P., Lundberg, G. D., & Rennie, D. (1998). Prevalence of articles with honorary authors and ghost authors in peer-reviewed medical journals. *Jama-Journal of the American Medical Association*, *280*(3), 222–224. http://doi.org/10.1001/jama.280.3.222

Giles, C. L., & Councill, I. G. (2004). Who gets acknowledged: Measuring scientific contributions through automatic acknowledgment indexing. *Proceedings of the*

*National Academy of Sciences of the United States of America*, *101*(51), 17599–604. http://doi.org/10.1073/pnas.0407743101

Henriksen, D. (2016). The rise in co-authorship in the social sciences (1980–2013). *Scientometrics*, *107*(2), 455-476. http://doi.org/10.1007/s11192-016-1849-x

Ilakovac, V., Fister, K., Marušić, M., & Marušić, A. (2007). Reliability of disclosure forms of authors' contributions. *Canadian Medical Association Journal*, *176*(1), 41–46. https://doi.org/10.1503/cmaj.060687

International Committee of Medical Journal Editors. (1997). Uniform Requirements for Manuscripts Submitted to Biomedical Journals. *New England Journal of Medicine*, *336*(4), 309–315.

International Committee of Medical Journal Editors. (2015, December). Recommendations. Retrieved from http://www.icmje.org/icmje-recommendations.pdf

Ivaniš, A., Hren, D., Marušić, M., & Marušić, A. (2011). Less Work, Less Respect: Authors' Perceived Importance of Research Contributions and Their Declared Contributions to Research Articles. *PLOS ONE*, *6*(6), e20206. https://doi.org/10.1371/journal.pone.0020206

Ivaniš, A., Hren, D., Sambunjak, D., Marušić, M., & Marušić, A. (2008). Quantification of Authors' Contributions and Eligibility for Authorship: Randomized Study in a General Medical Journal. *Journal of General Internal Medicine*, *23*(9), 1303–1310. https://doi.org/10.1007/s11606-008-0599-8

Katz, J. S., & Martin, B. R. (1997). What is research collaboration? Research Policy, 26(1), 1-18.

Larivière, V., Gingras, Y., Archambault, É. (2006). Canadian collaboration networks: A comparative analysis of the natural sciences, social sciences and the humanities. Scientometrics, 68(3): 519-533.

Larivière, V., Desrochers, N., Macaluso, B., Mongeon, P., Paul-Hus, A., & Sugimoto, C. R. (2016). Contributorship and division of labor in knowledge production. Social Studies of Science, 46(3), 417–435. https://doi.org/10.1177/0306312716650046

Laudel, G. (2002). What do we measure by co-authorships? *Research Evaluation*, *11*(1), 3-15. http://doi.org/10.3152/147154402781776961

Lundberg, J., Tomson, G., Lundkvist, I., Skar, J., & Brommels, M. (2006). Collaboration uncovered: Exploring the adequacy of measuring university-industry collaboration through co-authorship and funding. *Scientometrics*, *69*(3), 575–589. https://doi.org/10.1007/s11192-006-0170-5

Mackintosh, S. H. (1972). Acknowledgment patterns in sociology (Ph.D. Thesis). University of Oregon, Michigan.

Marušić, A., Bošnjak, L., & Jerončić, A. (2011). A Systematic Review of Research on the Meaning, Ethics and Practices of Authorship across Scholarly Disciplines. *PLoS ONE*, *6*(9), e23477. http://doi.org/10.1371/journal.pone.0023477

Marušić, A., Bates, T., Anić, A., & Marušić, M. (2006). How the structure of contribution disclosure statements affects validity of authorship: a randomized study in a general medical journal. *Current Medical Research and Opinion*, *22*(6), 1035–1044. https://doi.org/10.1185/030079906X104885

McCain, K. W. (1991). Communication, Competition, and Secrecy: The Production and Dissemination of Research-Related Information in Genetics. *Science, Technology & Human Values*, *16*(4), 491–516. http://doi.org/10.1177/016224399101600404

National Science Foundation. (2006). Science and Engineering Indicators. Chapter 5: Academic Research and Development. Data and Terminology. Retrieved from http://www.nsf.gov/statistics/seind06/c5/c5s3.htm#sb1

Paisley, W. J., & Parker, E. B. (1967). Scientific Information Exchange at an Interdisciplinary Behavioral Science Convention.

Paul-Hus, A., Desrochers, N., & Costas, R. (2016). Characterization, description, and considerations for the use of funding acknowledgement data in Web of Science. *Scientometrics*, *108*(1), 167-182. http://doi.org/10.1007/s11192-016-1953-y

Ponomariov, B., & Boardman, C. (2016). What is co-authorship? *Scientometrics*, 1–25. https://doi.org/10.1007/s11192-016-2127-7

Rennie, D., Yank, V., & Emanuel, L. (1997). When Authorship Fails: A Proposal to Make Contributors Accountable. *JAMA*, *278*(7), 579–585. http://doi.org/10.1001/jama.1997.03550070071041

Wislar, J. S., Flanagin, A., Fontanarosa, P. B., & DeAngelis, C. D. (2011). Honorary and ghost authorship in high impact biomedical journals: a cross sectional survey. *BMJ*, *343*, d6128. http://doi.org/10.1136/bmj.d6128

Wuchty, S., Jones, B. F., & Uzzi, B. (2007). The increasing dominance of teams in production of knowledge. *Science, 316*(5827), 1036–1039. doi:10.1126/science.1136099